\documentclass[10pt,prl,twocolumn,showpacs,floatfix,aps]{revtex4-1}   	
\usepackage{graphicx}			
\usepackage{amssymb}
\usepackage{natbib}

\begin{document}
\title{Excitation spectra in crystal plasticity}
\author{Markus Ovaska$^1$, Arttu Lehtinen$^1$, Mikko J. Alava$^1$, 
Lasse Laurson$^{1}$, Stefano Zapperi$^{1,2,3,4}$}
\affiliation{$^1$COMP Centre of Excellence,
Department of Applied Physics, Aalto University, P.O.Box 11100, 
FI-00076 Aalto, Espoo, Finland}
\affiliation{$^2$Center for Complexity and Biosystems, Department of 
Physics, University of Milano, via Celoria 16, 20133 Milano, Italy}
\affiliation{$^3$ISI Foundation, Via Alassio 11C, Torino, Italy}
\affiliation{$^4$ CNR - Consiglio Nazionale delle Ricerche, Istituto 
di Chimica della Materia Condensata e di Tecnologie per l'Energia, 
Via R. Cozzi 53, 20125 Milano, Italy}

\begin{abstract}
Plastically deforming crystals exhibit scale-free fluctuations that are 
similar to those observed in driven disordered elastic systems close 
to depinning, but the nature of the yielding critical point is
still debated. Here, we study the marginal stability of ensembles
of dislocations and compute their excitation spectrum in two and 
three dimensions. Our results show the presence of a singularity 
in the distribution of {\it excitation stresses}, i.e., the stress
needed to make a localized region unstable, that is remarkably similar
to the one measured in amorphous plasticity and spin glasses. 
These results allow us to understand recent observations of extended 
criticality in bursty crystal plasticity and explain how they originate 
from the presence of a pseudogap in the excitation spectrum.
\end{abstract}
\maketitle

Recent advances in crystal plasticity have revealed the
importance of collective dislocation dynamics in crystals 
\cite{zaiser2006scale,alava2014crackling,
ananthakrishna2007current}. The key phenomena include
power-law distributed strain bursts \cite{uchic2009plasticity,
dimiduk2006scale,ng2008stochastic,zaiser2008strain} and intermittent acoustic 
emission signals \cite{weiss2015mild,weiss1997acoustic,miguel2001intermittent},
originating from avalanches of dislocation activity.
The theoretical description of such crackling noise response in plasticity
has been debated, and ideas from depinning transitions 
\cite{ovaska2015quenched,PhysRevLett.109.095507}, jamming 
\cite{ispanovity2014avalanches,lehtinen2016glassy,PhysRevLett.89.165501}, 
and glassy dynamics \cite{bako2007dislocation} of the dislocation assembly 
have been brought up.

To resolve the question of the fundamental nature of bursty dislocation dynamics, 
it is central to understand how such activity bursts or
excitations are triggered by applied stresses. For amorphous plasticity, 
this has recently been addressed by considering the point-wise distribution 
$P(x)$ of the {\it local} distances $x$ in stress to a threshold above 
which a deformation burst is excited \cite{lin2014density,lin2014scaling,
liu2016driving}. In the $x \rightarrow 0$ limit, the form of $P(x)$  
encodes information about the nature of the dynamics exhibited 
by the system, and is connected with the properties 
of the ensuing crackling noise \cite{muller2015marginal}. One 
expects a power-law form $P(x) \propto x^{\theta}$, where the exponent 
$\theta$ characterizes the system. This approach stems from the classical 
depinning problem of elastic (a convex interaction kernel) interfaces in random 
media where $P(x)$ is flat for small $x$, so $\theta=0$. In models 
of amorphous plasticity as well as certain spin glass models singular behavior of 
$P(x)$ for $x \rightarrow 0$, or $\theta>0$ is found in the quasistatic limit 
\cite{lin2014density,lin2014scaling,liu2016driving,lin2015criticality,muller2015marginal}: 
a ``pseudogap'' in the excitation spectrum \cite{muller2015marginal}. 

It has been argued that the singular $P(x)$ is a consequence of the 
non-positive definite nature of the long-range interaction kernel mediating 
the collective deformation dynamics \cite{lin2014density,lin2014scaling}. 
This feature has been linked to the emergence of ``extended 
criticality'' in the crackling dynamics, when critical fluctuations 
take place over an extended range of control parameter values rather than only in the
proximity of a ``critical point'' \cite{muller2015marginal}. 
In the case of plastic deformation of a system with $N$ dislocations 
the argument states that if $P(x) \propto x^{\theta}$, the stress increment 
separating avalanches scales as $N^{-1/(1+\theta)}$, and hence the
number of events $N_{\text{a}}$ within a stress interval $\Delta \sigma \sim 1$ scales as 
$N_{\text{a}} \sim N^{1/(1+\theta)} \ll N$. However, these few events must be 
responsible for an extensive strain increment, such that the mean avalanche size 
scales as $N/N_{\text{a}} \sim N^{\theta/(1+\theta)}$, i.e., it diverges in the 
thermodynamic limit. This behavior has been 
observed in various glassy systems ranging from mean field spin glasses \cite{pazmandi1999self,
yan2015dynamics} to models of amorphous plasticity \cite{lin2015criticality}. 
Dislocation systems exhibit slow, glass-like dynamics, argued to originate from 
the frustrated dislocation interactions \cite{bako2007dislocation}. Together with 
recent results of extended criticality \cite{ispanovity2014avalanches,janicevic2015avalanches,
lehtinen2016glassy} this suggests the possibility of interesting, non-trivial 
excitation spectra. 

In this letter, we report an extensive study of the excitation spectra in 
crystal plasticity by considering 2D and 3D discrete dislocation dynamics (DDD) 
systems. We perform simulations with and without the presence of quenched pinning
centers (e.g. solute atoms). The key findings are that the distributions 
$P(x)$ are singular in pure systems and 
develop a small $x$ cut-off with quenched disorder. In all cases 
considered the exponent $\theta$ is found to depend on dimension, the stress rate 
of the local perturbation, the presence or absence of quenched pinning, and 
whether the other dislocations not directly subject to the perturbation are 
allowed to move or not.

\begin{figure}[t!]
\centering
\includegraphics[width=10 cm]{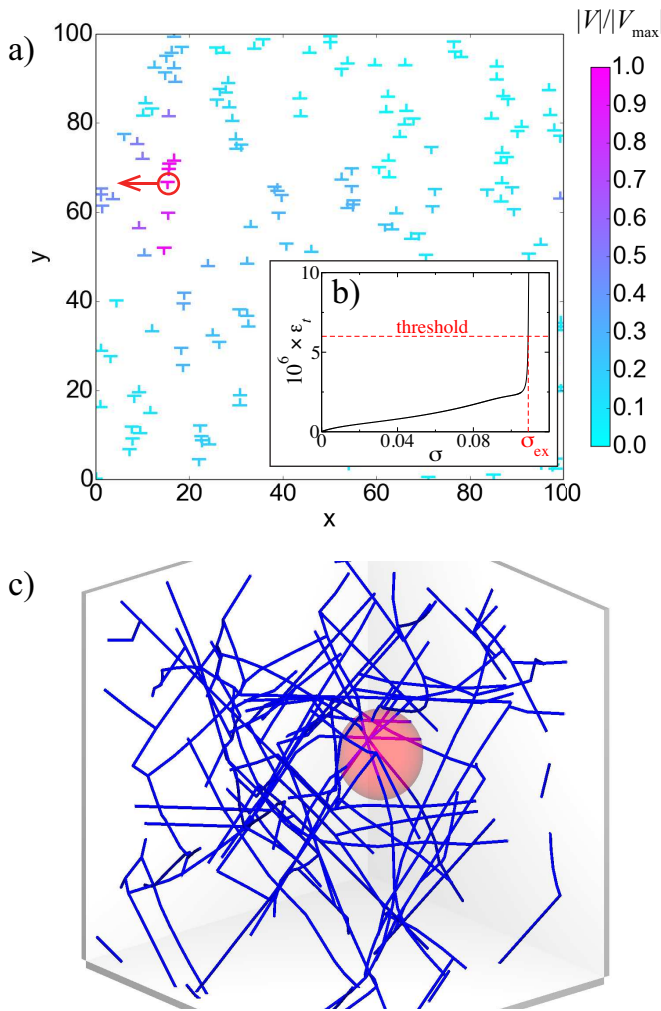}
\caption{Idea of the 2D and 3D DDD simulations used to probe the
stability of dislocation assemblies against local stress perturbations. a) shows an example
from a 2D DDD simulation, where a single dislocation (surrounded by a small circle) is 
subject to a linearly increasing local external stress $\sigma$ (indicated by the arrow),
with the color code corresponding to the instantaneous velocity of the dislocations, showing
that the dislocation activity tends to spread beyond the dislocation directly subject to
the perturbation. b) displays an example of the strain rate vs stress graph, along with an
illustration of the procedure to define an excitation by thresholding the strain rate signal.
c) shows an example of the 3D dislocation system, with the red sphere corresponding to the
location where the local external stress $\sigma$ is being applied to excite dislocation 
activity.}
\label{Fig:1}
\end{figure}

The 2D and 3D DDD models we consider are similar to the standard ones discussed in the literature. The 2D
case is studied in more depth as the 3D one is intrinsically hard numerically. The starting point in both cases
is a relaxed, zero-stress state, which we expect to be in the extended criticality regime of the pure
dislocation systems \cite{ispanovity2014avalanches,janicevic2015avalanches,lehtinen2016glassy}.
The 2D DDD model represents a cross section ($xy$ plane) of a single crystal, with a single 
slip geometry, and straight parallel edge dislocations along the $z$ axis. The $N_{\text{d}}$ edge 
dislocations glide along directions parallel to their Burgers vectors ${\bf b} = \pm b {\bf u}_x$, 
where $b$ is the magnitude and ${\bf u}_x$ is the unit vector along the $x$ axis. Equal numbers 
of dislocations with positive and negative Burgers vectors are assumed, and dislocation climb 
is not considered. The dislocations interact with each other through their long-range stress 
fields $\sigma_{xy}=Dbx(x^2-y^2)/(x^2+y^2)^2$, where $D=\mu/2\pi(1-\nu)$, $\mu$ is the 
shear modulus, and $\nu$ the Poisson ratio. Dislocation annihilation taking place in real crystals
is modelled by removing dislocation pairs with opposite Burgers vectors from the system if their mutual 
distance is smaller than $b$. In addition, to test the effect of quenched
disorder on the excitation spectrum, we perform a set of simulations including also $N_{\text{s}}$
randomly positioned immobile pinning centers (``solute atoms'') interacting with the dislocations
via short range interactions, employing a regularized interaction energy $U=As_n \sin \theta/r[1-\exp 
(-(k^2/a^2)r^2)]$, with $k=1.65$, $a$ the atomic distance, $s_n$ the sign of the Burgers vector of
the dislocation, and $A=(1+\nu)\mu b \Delta V/3\pi(1-\nu)$ an interaction strength parameter, 
with $\Delta V$ the misfit area \cite{wang1990non}. These pinning centers generate a short-range 
correlated random pinning field (with a correlation length $\xi \approx (N_{\text{s}}/L^2)^{-1/2}$
and strength proportional to $A$) interfering with the dislocation motion \cite{ovaska2015quenched}. 
Dislocation dynamics is taken to be overdamped, such that the dislocation velocity is proportional to the
Peach-Koehler force due to the total stress (with contributions from dislocation interactions,
the random pinning field, as well as from the external stress $\sigma_{\text{ext}}$ when present) 
acting on it.

The 3D DDD simulations are performed using the ParaDis code \cite{arsenlis2007enabling}, considering the 
FCC crystal structure with material parameters of Al (shear modulus $G$ = 26 GPa, Poisson ratio 0.35, 
Young modulus 70.2 GPa, Burgers vector $b$ = 2.863 $\times$ $10^{-10}$ m, and dislocation 
mobility $10^4$ Pa$^{-1}$ s$^{−1}$; for simplicity, both edge and screw segments are taken to 
have the same mobility). The line dislocations are modeled using a nodal 
discretization scheme where dislocation lines are represented by nodal points connected to 
their neighbors by straight segments. Changes in dislocation geometry are modeled by 
adding and removing these nodal points. The total stress acting on a node consists of the 
external part, resulting from the deformation of the whole crystal, and of the internal, 
anisotropic stress fields generated by the other dislocations within the crystal. This leads
to a Peach-Koehler force moving the discretization nodes according to a material-specific 
mobility function which relates the total forces experienced by dislocations to their 
velocities, encoding also the constraints on dislocation motion due to crystal structure. 
Forces between segments of nearby nodes and self-interaction of dislocations are calculated 
with explicit line integrals, while the far-field forces are computed from the coarse-grained 
dislocation structure using a multipole expansion. In addition, near the dislocation core, 
local interactions, such as junction formation, annihilation, etc., are introduced 
phenomenologically with input from smaller scale simulation methods (e.g., MD) and 
experimental results. A trapezoidal integrator is then used to solve the equation of 
motion for the discretization nodes. We consider periodic boundary conditions for both 
2D and 3D simulations. In each case, a random dislocation configuration (with $N_{\text{d}}=400$ in 
2D and 40 dislocation lines in 3D) is first let to relax in the absence of applied stresses; 
after the relaxation, ~200 dislocations remain in 2D after the annihilations during relaxation, 
and the initially straight dislocation lines in 3D develop some corvature and form junctions. 
We then proceed to study the stability of these relaxed dislocation configurations.

\begin{figure}[t!]
\centering
\includegraphics[width=8.5 cm]{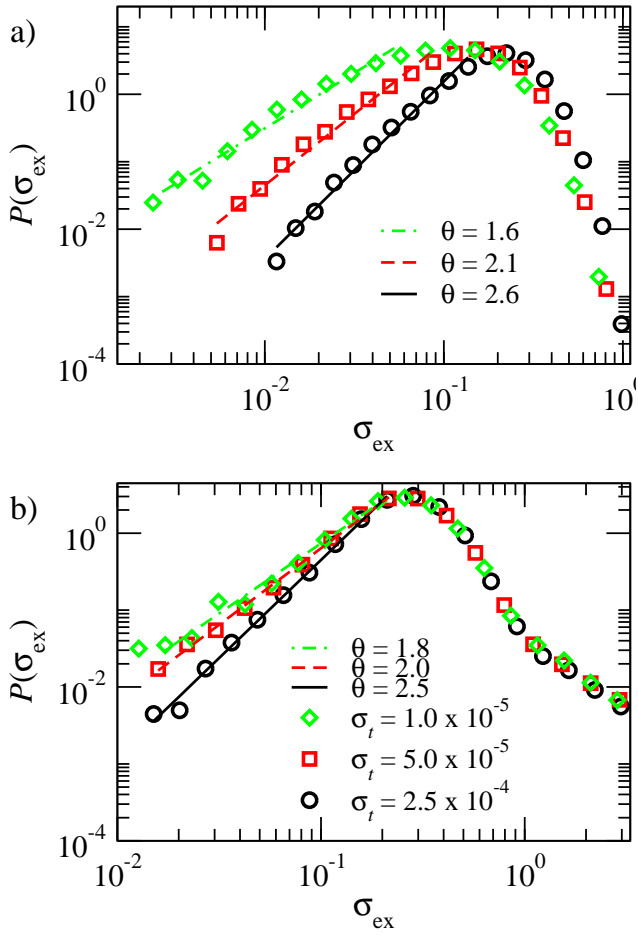}
\caption{Excitation spectra $P(\sigma_{\text{ex}})$ from
2D DDD simulations for three different stress rates $\sigma_t$ (indicated in the legend)
for pure dislocation systems. a) shows $P(\sigma_{\text{ex}})$ for systems where all the
dislocations are allowed to move, while in b) the corresponding data for systems where
only the dislocation subject to the local stress perturbation is mobile is displayed.
All distributions are singular ($\theta >0$) for small $\sigma_{\text{ex}}$, with the lines
corresponding to fits of the form of $P(\sigma_{\text{ex}}) \propto \sigma_{\text{ex}}^{\theta}$.}
\label{Fig:2}
\end{figure}

\begin{figure}[t!]
\centering
\includegraphics[width=8.5 cm]{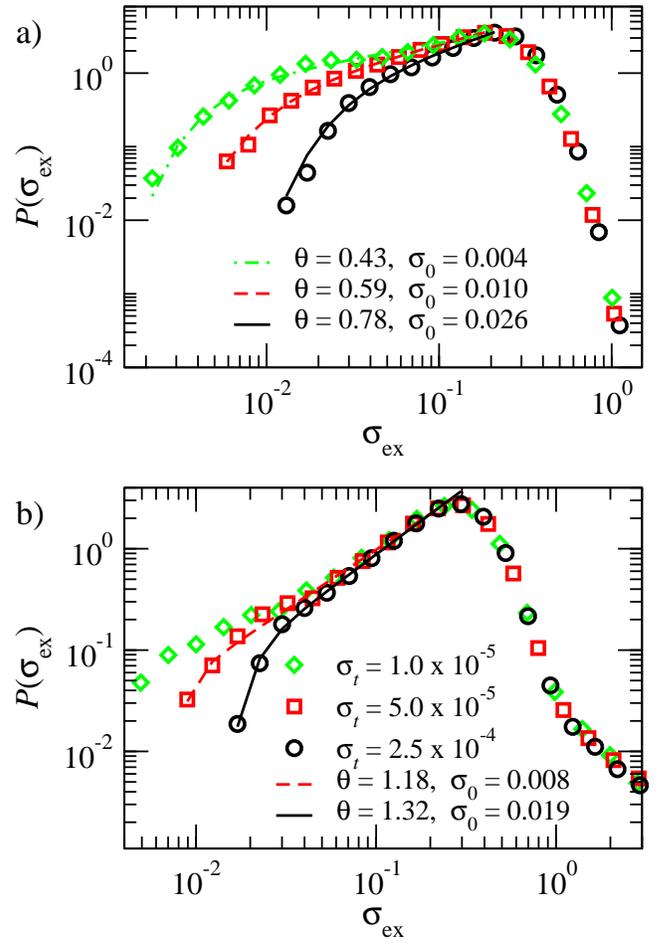}
\caption{Excitation spectra $P(\sigma_{\text{ex}})$ from
2D DDD simulations for three different $\sigma_t$ for dislocation
systems containing quenched disorder of ``intermediate'' strength, inducing depinning-like
dislocation dynamics. In a), all the dislocations are mobile, while in b), only the
dislocation subject to $\sigma$ is allowed to move. The distributions are
singular for small $\sigma_{\text{ex}}$, down to a small-$\sigma_{\text{ex}}$ cutoff at
$\sigma_0$ due to quenched disorder. Lines are fits of the form of $P(\sigma_{\text{ex}}) 
\propto \sigma_{\text{ex}}^{\theta} \exp [-(\sigma_0/\sigma_{\text{ex}})^{\alpha}]$.}
\label{Fig:3}
\end{figure}

To create excitations in dislocation systems by local stress perturbations,
we ramp up a {\it local} external stress $\sigma$ from zero at a constant rate $\sigma_t$,
and monitor the response of the system by considering the resulting time-dependent
{\it global} strain rate $\epsilon_t$. 
In the 2D simulations, we go through each individual 
dislocation one by one, and apply a positive scalar $\sigma=\sigma_{xy}$ only to that 
[see Fig. \ref{Fig:1}a)]. In the 3D case, we choose to apply a local tensile stress within a 
spherical region, with an example shown in Fig. \ref{Fig:1}c) as the red sphere; we consider spheres
with radii comparable to the average dislocation spacing $1/\sqrt{\rho}$, where $\rho$
is the dislocation density, and go through all such non-ovelapping spheres in each system. 
This particular protocol is mandated by the procedure by which
line dislocations are treated in a DDD code such as ParaDis. For both 2D and 3D, 
an excitation is defined when
$\epsilon_t$ first exceeds the threshold at a stress $\sigma=\sigma_{\text{ex}}$,
see Fig. \ref{Fig:1}b). The results are averaged over both different relaxed 
dislocation configurations (500 and 100 in 2D and 3D, respectively) as well as over different 
excitation locations in each configuration, i.e., ~200 dislocations and ~400 spheres per system, 
resulting in approx. $10^5$ and $4 \times 10^4$ local stress-strain rate curves in 2D and 3D, 
respectively. In 3D, not all of these lead to an excitation due to the tensile applied 
stress not always producing the relevant resolved shear stress(es) within the excitation volume.

In the 2D case, the $P(\sigma_{\text{ex}})$s
are summarized in Figs. \ref{Fig:2} and \ref{Fig:3}. The $P(\sigma_{\text{ex}})$s
for three different stress rates $\sigma_t$ for the ``pure'' dislocation system are shown
in Fig. \ref{Fig:2}a). For small $\sigma_{\text{ex}}$, a clear power-law form 
$P(\sigma_{\text{ex}}) \propto \sigma_{\text{ex}}^{\theta}$ is observed, indicating singular
behavior of $P(\sigma_{\text{ex}})$ for $\sigma_{\text{ex}} \rightarrow 0$, with the 
exponent $\theta$ depending on the stress rate; $\theta$ evolves from $\theta \approx 2.6$ 
to $\theta \approx 1.6$ as the stress rate is lowered from $\sigma_t = 2.5 \times 10^{-4}$
to $1 \times 10^{-5}$. Notice that rate-dependent $\theta$-exponents have been observed also 
in elasto-plastic models for amorphous materials \cite{liu2016driving}. It is worth pointing 
out that in the case considered in Fig. \ref{Fig:2}a), all the dislocations are able to respond 
to the local stress perturbation via the long-range dislocation-dislocation interactions, 
and thus it sometimes happens that the dislocation activity induced by the local stress 
is not localized in the immediate vicinity of the dislocation subject to the perturbation;
this is illustrated by Movie 1 (Supplemental Material \cite{SM}). For comparison,
Fig. \ref{Fig:2}b) shows the corresponding $P(\sigma_{\text{ex}})$s in a system where
all the other dislocations, except for the one directly subject to the stress perturbation,
are kept fixed, in analogy to the procedure employed when considering various lattice models
of amorphous plasticity \cite{lin2014scaling}. While the $P(\sigma_{\text{ex}})$s in Fig. \ref{Fig:2}b) 
look qualitatively somewhat different from those shown in Fig. \ref{Fig:2}a), showing in particular 
a weaker overall dependence on the stress rate $\sigma_t$, the $\sigma_t$-dependent 
exponent $\theta$ nevertheless assumes similar values in both cases: even the single-dislocation 
excitations display singular characteristics. In all these cases the spectra show that creating 
excitations becomes harder with higher stress rates: a larger $\sigma_{\text{ex}}$ is typically
required for a larger $\sigma_t$.

In presence of quenched disorder, the behavior described above for the pure dislocation 
systems should change. We show in Fig. \ref{Fig:3}a) the $P(\sigma_{\text{ex}})$s for 
a system with a quenched pinning field generated by randomly distributed pinning 
centers of ``intermediate'' strength, with $A=0.1$ and $N_{\text{s}}/L^2=0.8 b^{-2}$ chosen 
to transform the glass-like jamming scenario of pure dislocation systems into a depinning-like 
problem, but the disorder is not so strong that it would eliminate the scale-free nature of 
the avalanches (see Ref. \cite{ovaska2015quenched} for details); we consider the same 
three stress rates as above to apply the local perturbation. One may make two main observations: 
(i) the exponent $\theta$ assumes 
lower stress rate dependent values as compared to the pure dislocation system, and 
(ii) the power laws exhibited by the $P(\sigma_{\text{ex}})$s appear to have 
low-$\sigma_{\text{ex}}$ cutoffs - hence we fit the data with the function $P(\sigma_{\text{ex}}) \propto
\sigma_{\text{ex}}^{\theta} \exp [-(\sigma_0/\sigma_{\text{ex}})^{\alpha}]$, where $\sigma_0$ is the
cutoff stress scale. These cutoffs are related to the finite stress needed to push 
the dislocation out of the local potential energy minimum due to the quenched pinning field:
thus, we excpect $\sigma_0$ to increase with increasing disorder strength. 
For a given disorder strength, $\sigma_0$ is found to increase with increasing $\sigma_t$.
Notice that no such feature is observable in the pure dislocation system, where dislocation
activity may in some instances be triggered with a very small $\sigma_{\text{ex}}$.
Fig. \ref{Fig:3}b) shows the results from the 2D system with pinning and all the
other dislocations fixed; again, a small-$\sigma_{\text{ex}}$ cutoff can be observed,
and the distributions show less sensitivity to $\sigma_t$ than in the case where
all dislocations are able to respond to the perturbation. 

\begin{figure}[t!]
\centering
\includegraphics[width=8.5 cm]{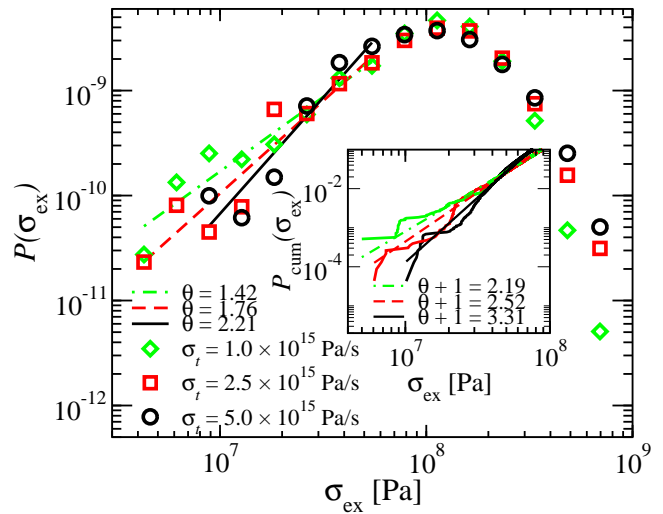}
\caption{Excitation spectra $P(\sigma_{\text{ex}})$ from
3D DDD simulations for three different $\sigma_t$. All dislocations are
allowed to move, and there is no quenched disorder. The $\sigma_t$ dependent $\theta$-exponents
are similar to those found in the corresponding 2D DDD model, and may be approximately
recovered also by considering the cumulative distribution function 
$P_{\text{cum}}(\sigma_{\text{ex}})$ (inset).}
\label{Fig:4}
\end{figure}

Finally, we consider the excitations in the 3D DDD simulations, considering for 
simplicity and due to the high computational cost of the 3D simulations
only the case where all dislocations are mobile, and no quenched disorder is present;
see Movie 2 in the Supplemental Material \cite{SM} for an example of the excitation process.
Fig. \ref{Fig:4} shows that the results are qualitatively similar to those found
above for the 2D system: the distributions $P(\sigma_{\text{ex}})$ are singular
at $\sigma_{\text{ex}} \rightarrow 0$, and the exponent $\theta$ exhibits a similar
dependence on $\sigma_t$ as that shown in Fig. \ref{Fig:3} for the corresponding
2D system, with $\theta$ varying between $\sim 2.2$ and $\sim 1.4$ for the data shown
in Fig. \ref{Fig:4}. The inset shows the corresponding cumulative distribution function 
$P_{\text{cum}}(\sigma_{\text{ex}})$ for small $\sigma_{\text{ex}}$, exhibiting scaling
with an exponent $\theta+1$, with the $\theta$-values in reasonable agreement with
those estimated from the $P(\sigma_{\text{ex}})$s. 
Considering different threshold values for the strain rate
to define the excitations leads to slightly threshold-dependent exponents in the 3D
case (not shown), but the qualitative picture remains the same for a range of
threshold values. Thus, a ``pseudogap'' appears to be present in the excitation 
spectrum also in 3D dislocation systems.

To summarize, we have performed both 2D and 3D DDD simulations to establish that the 
excitation spectra in crystal plasticity exhibit singular behavior at 
$\sigma_{\text{ex}} \rightarrow 0$, something that is at odds with the known
behavior of elastic manifolds in random media exhibiting a depinning transition.
As this behavior persists when moving from 2D systems to 3D ones, the root cause  
is shared by both systems. The likely cause is the anisotropic, non-positive definite 
interactions between dislocations, in analogy to the quadrupolar Eshelby-type stress 
fields argued to be responsible for the similar behavior found recently for amorphous 
plasticity \cite{lin2014density,lin2014scaling}. This then persists upon coarse-graining, 
and manifests itself in the presence of critical-like fluctuations or plastic avalanches 
even at negligible external stresses \cite{ispanovity2014avalanches,janicevic2015avalanches,
lehtinen2016glassy}. The presence of a frozen impurity field on the other hand leads to a 
finite minimum for the excitation stress $\sigma_{\text{ex}}$, in agreement with the idea 
of a presence of a true critical point or yield stress \cite{ovaska2015quenched} 
instead of the extended criticality scenario of pure dislocation systems. 
Finally, we point out that 2D colloidal crystals \cite{pertsinidis2005video} might 
provide an interesting experimental system to test our results.

\begin{acknowledgments}
This work has been supported by the Academy of Finland through its Centres of 
Excellence Programme (2012-2017) under project no. 251748, an Academy Research 
Fellowship (LL, project no. 268302), and the FiDiPro program (SZ, project no. 
13282993). We acknowledge the computational resources provided by the 
Aalto University School of Science ``Science-IT'' project, as well as those 
provided by CSC (Finland). SZ is supported by the ERC Advanced Grant no. 291002 SIZEFFECTS.
\end{acknowledgments}

\bibliographystyle{apsrev4-1} 
\bibliography{literature}

\end{document}